\documentclass[bjps]{imsart}

\RequirePackage{amsthm,amsmath,amsfonts,amssymb}
\RequirePackage[authoryear]{natbib}
\startlocaldefs
\numberwithin{equation}{section}
\theoremstyle{plain}
\usepackage{graphicx}
\usepackage{amsbsy}
\usepackage{paralist}
\usepackage{bm}
\usepackage{booktabs}
\usepackage{wrapfig}

\usepackage{pgfplots}

\pgfplotsset{compat=1.9}

\usepackage{subcaption}

\usepackage{mathtools}
\usepackage{cases}

\usepackage{algorithm}
\usepackage[noend]{algpseudocode}

\usepackage{tcolorbox}

\RequirePackage[hyphens]{url}
\usepackage{color}
\usepackage{hyperref}
\definecolor{darkblue}{rgb}{0.0,0.0,0.3}
\hypersetup{colorlinks,breaklinks,linkcolor=darkblue,urlcolor=darkblue,
            anchorcolor=darkblue,citecolor=blue}
\renewcommand{\d}{{\rm d}}  % for derivatives
\newcommand{\e}{{\rm e}} % for exponentials

\newcommand{\x}{{\bf x}}

\newcommand{\E}{{\mathbb E}}

\newcommand{\X}{{\bf X}}

\newcommand{\btheta}{\boldsymbol{\theta}}

\newcommand{\1}{{\bf 1}}

\newcommand{\ben}{\begin{enumerate}}
\newcommand{\een}{\end{enumerate}}
\newcommand{\beq}{\begin{equation}}
\newcommand{\eeq}{\end{equation}}
\newcommand{\bde}{\begin{description}}
\newcommand{\ede}{\end{description}}

\newcommand{\half}{\frac{1}{2}}

\newcommand{\abs}[1]{\lvert#1\rvert}

\newcommand{\NormRV}{\mathcal{N}}

\newcommand{\UnifRV}{\mathcal{U}}

\newcommand{\BetaRV}{\mathcal{B}eta}

\newcommand{\iidd}{\stackrel{\mathrm{D}}{=}}

\newcommand{\defeq}{\operatorname{:=}}

\newtheorem{theorem}{Theorem}

\newtheorem{lemma}[theorem]{Lemma}

\newtheorem{remark}[theorem]{Remark}

\usepackage{booktabs,array}

\newcount\rowc

\endlocaldefs

\begin{document}

\begin{frontmatter}

\title{Quantile Importance Sampling}
%\title{A sample article title with some additional note\thanksref{T1}}
\runtitle{Quantile Importance Sampling}
%\thankstext{T1}{A sample of additional note to the title.}

\begin{aug}
%% \ead[label=o1,orcid]{0000-0000-0000-0000} %%
%%%%%%%%%%%%%%%%%%%%%%%%%%%%%%%%%%%%%%%%%%%%%%%
\author[A]{\inits{Dr.}\fnms{Jyotishka}~\snm{Datta}\ead[label=e1]{jyotishka@vt.edu}},
\author[B]{\inits{Dr.}\fnms{Nicholas G.}~\snm{Polson}\ead[label=e2]{ngp@chicagobooth.edu}}

%%%%%%%%%%%%%%%%%%%%%%%%%%%%%%%%%%%%%%%%%%%%%%
%% Addresses                                %%
%%%%%%%%%%%%%%%%%%%%%%%%%%%%%%%%%%%%%%%%%%%%%%
\address[A]{Department of Statistics, Virginia Tech, Blacksburg, VA \printead[presep={.\ }]{e1}}

\address[B]{Booth School of Business, University of Chicago, Chicago, IL \printead[presep={.\ }]{e2}}
\end{aug}

\begin{abstract}
\noindent In Bayesian inference, the approximation of integrals of the form $\psi = \mathbb{E}_{F}{l(X)} = \int_{\chi} l(\mathbf{x}) dF(\mathbf{x})$ is a fundamental challenge. Such integrals are crucial for evidence estimation, which is important for various purposes, including model selection and numerical analysis. The existing strategies for evidence estimation are classified into four categories: deterministic approximation, density estimation, importance sampling, and vertical representation \citep{llorente2020marginal}. In this paper, we show that the Riemann sum estimator due to \citet{yakowitz1978weighted} can be used in the context of nested sampling \citep{skilling2006nested} to achieve a $O(n^{-4})$ rate of convergence, faster than the usual Ergodic Central Limit Theorem, under certain regularity conditions. We provide a brief overview of the literature on the Riemann sum estimators and the nested sampling algorithm and its connections to vertical likelihood Monte Carlo. We provide theoretical and numerical arguments to show how merging these two ideas may result in improved and more robust estimators for evidence estimation, especially in higher dimensional spaces. We also briefly discuss the idea of simulating the Lorenz curve that avoids the problem of intractable $\Lambda$ functions, essential for the vertical representation and nested sampling. 
\end{abstract}


\end{frontmatter}
%%%%%%%%%%%%%%%%%%%%%%%%%%%%%%%%%%%%%%%%%%%%%%
%%%% Main text entry area:
\section{Introduction}\label{sec:intro}

In Bayesian inference, a fundamental challenge is the approximation of integrals of the form $\psi = \E_{F}{L(\btheta)} = \int_{\chi} L(\btheta) dF(\btheta)$, where $L(\btheta)$ is a measurable function of interest and $F(\cdot)$ is a finite measure, such as a probability. Such integrals are crucial for \textit{evidence} estimation, that is, the approximation of the quantity: 
\[
Z = \int_{\Theta} L(\x \mid \btheta) \pi(\btheta) d\btheta,
\]
where $L(\btheta) = L(\x \mid \btheta)$ is the likelihood with observed data $\x$, $\btheta \in \mathbb{R}^p$ is the parameter of interest and $\pi(\btheta)$ is the chosen prior over $\btheta$. The \textit{evidence} is also called marginal likelihood and the normalizing constant as one can write the posterior distribution of $\btheta$ as $P(\btheta \mid \X) = L(\x \mid \btheta) \pi(\btheta)/ Z$, ensuring $\int_{\mathbb{R}^p} P(\btheta \mid \x) d\btheta = 1$. Evidence estimation is significant for various purposes, including the calculation of Bayes factors for model comparison and numerical analysis, as well as in statistical mechanics. The field of evidence estimation has been the subject of extensive research and is well-documented in the literature, as summarized in \citet{llorente2020marginal}. In \citet{llorente2020marginal}, the existing strategies for evidence estimation are classified into four categories: deterministic approximation, density estimation, importance sampling, and vertical representation, which encompasses nested sampling \citep{skilling2006nested}.  

Monte Carlo methods have a long and rich history starting with Stan Ulam and John von Neumann in 1946 when Stan Ulam, while recuperating from an illness, had the vision that the probability of winning in a game of solitaire can be estimated from the proportions of winning hands from repeated simulations of the game, avoiding the combinatorial intractability. Ulam and Neumann realized that this can be extended to other difficult problems in mathematical physics like neutron diffusion \citep{eckhardt1987stan}. The subsequent work by many scientists at Los Alamos National Laboratory culminated in the foundation of Monte Carlo methods, landmarked by the milestone papers: \citet{metropolis1949monte, metropolis1953equation}, ushering in a new era of simulation-based methods aided by the explosive growth in computing speed and availability. We refer the interested readers to extensive reviews of Bayes computing in \citep{liu2001monte, robert1999monte, martin2023computing}, and the references therein. 

The na\"ive Monte Carlo estimator for evidence as well as the usual importance sampling method for estimating $\psi$ involves computing the empirical average of a sample $(x_1, \ldots, x_n)$ drawn from either the target density $f$ or a proposal density $g$:
\begin{equation}
\hat{\psi}_{IS} = \frac{1}{n} \sum_{i=1}^{n} \frac{L(x_i) f(x_i)}{g(x_i)} \label{eq:is}.
\end{equation}
If the target density is easy to sample from, \textit{i.e.}, if we can take $f = g$, the above expression reduces to the simple empirical average $\hat{\psi} = n^{-1} \sum_{i=1}^{n} L(x_i)$, or the na\"ive Monte Carlo. Using the Law of Large Numbers, one can show that the $\hat{\psi}_{IS}$ converges to the true value $\psi$ at a $O(n^{-1})$ rate. The density function $g(\cdot)$ is selected such that the weights $w(x_i)$ are approximately constant and various alternative schemes have been proposed to increase their efficiency. Examples include the harmonic mean (HM) estimator \citep{newton1994approximate}, criticized for its possible infinite variance \citep{neal1994}, the stabilized HM estimator \citep{raftery2006estimating}, or \citet{firth2011improved}'s improved estimator based on difference estimators. \citet{hesterberg1988advances, hesterberg1995weighted} and \citet{firth2011improved} introduced and refined the ratio estimator for importance sampling: $\sum_{i=1}^{n} l(x_i) w(x_i)/\sum_{i=1}^{n}w(x_i)$, where the weight $w(\cdot)$ can be chosen to move the sampler away from the low probability regions.\\

\noindent \textbf{\textcolor{darkblue}{Riemann Sums:}} The $O(n^{-1})$ rate can be improved greatly by using the Riemann sums, \textit{i.e.}, the trapezoidal rule where one can utilize ordered samples from the target distribution. For integrals over the unit interval $\int_0^1 L(x)dx$, \textit{i.e.}, in the case of uniform $f(\cdot)$, \citet{yakowitz1978weighted} showed that the Riemann estimator given by \eqref{eq:yakowitz} attains a $O(n^{-4})$ rate of convergence. This is a faster convergence rate than the Central Limit Theorem, and can be proved using the higher-order moments of the Dirichlet law and a well-known bound on the error of trapezoid rule. 
\begin{equation}
   \hat{\psi}_{Y} = \sum_{i=0}^{n-1}\frac{L(u_{(i)}) + L(u_{(i+1)})}{2} (u_{(i+1)} - u_{(i)}), \; \text{for} \; \psi = \int_0^1 L(x)dx, \label{eq:yakowitz}
\end{equation}
where $u_{(i)}$ denotes the $i^{th}$ order statistics for a sample $u_1, \ldots, u_n \sim \UnifRV(0,1)$. We shall now point the readers to an unexplored connection between the Yakowitz estimator \eqref{eq:yakowitz} for integrals of type $\int_0^1 L(x)dx$ and the Lorenz identity, a key idea that appears in nested sampling and in more general vertical likelihood methods \citet{skilling2006nested, chopin2010properties, polson2014vertical}. We discuss the details in section \ref{sec:yakowitz}. 

\vspace{0.1in} 
\noindent \textbf{\textcolor{darkblue}{Vertical Likelihood:}} The central idea behind Vertical Likelihood Monte Carlo \citep{polson2014vertical} is to represent a $p$-dimensional integral $\psi = \int_{\chi} L(\x) dF(\x) = \int_{\chi} L(\x) f(\x) \d\x$ as a one-dimensional integral using a latent variable parameter expansion. To do this, we compute $Z$ in two equivalent ways: via the tail integral of the survival function of $Y=L(\X)$, or equivalently, via the area under the Lorenz curve.

Suppose that $\X \sim P(\x)$ and $Y \equiv L(\X)$, called the \textit{likelihood ordinate}. Let $F_Y(y) = \mathbb{P}\{L(\X) \leq y\}$ be the cumulative distribution function of $Y$.  Now define the upper cumulant, or the survival function for the likelihood ordinate $Y \equiv L(\X)$: 
\begin{equation}
Z(y) = 1 - F_Y(y) = \int_{L(\x) > y} \d P(\x). \label{eqn:Zu}
\end{equation}
The function $Z(y) \in [0,1]$ is also called the \textit{volume variable} in \citet{ashton2022nested}, as it is the volume enclosed by the likelihood contour. Evidently, $Z(y)$ has domain $\mathbb{R}^+$ and range $[0,1]$ and is non-increasing in $y$. Let us denote the pseudo-inverse of $Z(y)$, by $\Lambda(s)$, defined as:
\begin{equation}
\Lambda(s) = \sup\{y: Z(y) > s\} \;\label{eqn:LambdaZ},
\end{equation}
which, like $Z(y)$, is also non-increasing. Using these two key functions, we can write the evidence $Z$ (for any $p$ with $\x \in \mathbb{R}^p$) as:
\begin{equation}
    Z = \int_0^{\infty} Z(y) \d y = \int_0^1 \Lambda(s) \d s \label{eq:master}.
\end{equation}
It also follows from the definitions of $\Lambda(s)$ and $Z(y)$ that $Z(L(\X)) \iidd \UnifRV(0,1)$ and consequently, $\Lambda(U) \iidd L(\X)$, \textit{i.e.,} the implied distribution of the likelihood ordinates are the same as the Lorenz curve $\Lambda(\cdot)$ evaluated at uniform grid points. This Lorenz identity along with equation \eqref{eq:master} is the key result that lets one evaluate any higher dimensional integral as a univariate integral over $[0,1]$, and use more efficient methods of approximation. Note that simple na\"ive Monte Carlo approaches will likely fail in higher dimensions because the survival function $Z(s)$ changes much more rapidly in its domain compared to $L(\x \mid \btheta) \pi(\btheta)$: necessitating the choice of `good grid points' for evaluating the Riemann sum. To quickly see why \eqref{eq:master} holds true, consider the following result: 
\begin{equation}
Z = \int_{\chi} L(x) \d P(x) =  \int_{\chi} \int_0^{\infty}  \mathbb{I}\{y < L(x)\}  \d y \d P(x) = \int_0^{\infty} Z(y) \d y  \label{eqn:Zin1D} \, 
\end{equation}
Intuitively, $\Lambda(s)$ gives the value $y$ such that $s$ is the fraction of prior draws with likelihood values larger than $y$. Now, exploiting the fact that $\{s < Z(y)\}$ if and only if $\{y < \Lambda(s)\}$, we can write: 
\begin{equation}
Z =  \int_0^{\infty} \int_0^1 \mathbb{I}\{ s < Z(y) \} \d s \d y 
=  \int_0^{\infty} \int_0^1 \mathbb{I}\{ y < \Lambda(s) \} \d s \d y =   \int_0^1 \Lambda(s) \d s \label{eqn:Zin1D2} \, .
\end{equation}

Critically, we do not have to assume that either $F^{-1}(s)$ or $\Lambda(s)$ \eqref{eqn:LambdaZ} are available in closed form, as we can find an unbiased estimate of this by simulating the Lorenz curve, as discussed in section \ref{sec:lorenz}, following \citet{chopin2010properties} or \cite{polson2014vertical}. 

This identity \eqref{eqn:Zin1D2} is exactly in the form of univariate integral in \eqref{eq:yakowitz} considered in \citep{yakowitz1978weighted} as $s$ is uniform on $[0,1]$. Hence, we can expect to obtain a huge improvement in the convergence rate to the order of $O(n^{-4})$, provided that the integrand $\Lambda(\cdot)$ has continuous, second derivative. We discuss the Yakowitz estimator in details in the next section \ref{sec:yakowitz}. The other identity \eqref{eqn:Zin1D} is important in nested sampling \citep{skilling2006nested}.

\vspace{0.1in} 
\noindent \textbf{\textcolor{darkblue}{Nested Sampling:}} \citet{polson2014vertical} shows the connection or equivalence between the Vertical Likelihood idea and the popular Nested Sampling approach by John Skilling \citep{skilling2006nested}, that also provides a way to transform the problem of high-dimensional integration into a sequence of lower-dimensional integrals. Nested Sampling (NS) proceeds by iteratively sampling a set of `live' points from the prior distribution subject to a constraint that the likelihood of the samples is greater than some threshold value, while discarding the lowest likelihood samples at each iteration. 

As the threshold is gradually lowered, the samples move towards regions of higher likelihood and the prior volume enclosing them shrinks, roughly in the order of a $\BetaRV(n,1)$ random variate for $n$ live points. At each step of the process, the algorithm discards the lowest likelihood samples and replaces them with new ones drawn from the prior distribution subject to the same constraint. This continues until the threshold reaches a very small value, at which point the remaining samples are concentrated in the high-likelihood regions of the posterior distribution. The evidence of the model can be estimated from the discarded samples, and the posterior distribution can be approximated using the remaining samples. We present the NS algorithm in appendix section \ref{sec:nested}. \\

% Lorenz curve of the likelihood ordinate $Y \equiv L(X)$ where $X \sim p(x)$. 
% The Lorenz curve, $\mathcal{L}(X)$ is defined in terms of its CDF, $F_X(x)$, as 
% \begin{align*}
%   \mathcal{L}(u) & = \frac{1}{Z} \int_0^u F_X^{-1} ( s ) d s \; \; \text{ where } \; u \in [0,1] \\
%   Z & = \mathbb{E} ( X) = \int_{ \mathcal{X} } L(x) p(dx ) \; .
% \end{align*}
% One feature of a Lorenz curve is that it provides a way to evaluate: 
% \[
% \mathbb{E}(X) = \int_{\mathcal{X}} L(x)p(dx) = \int_0^1 F_X^{-1}(s) ds
% \] 
% Our aim is to summarize the literature on this class of estimators while showing how marrying these two ideas may result in faster convergence for evidence estimation.

In this paper, we explore the vertical representation of evidence and articulate how the Riemann sum estimator, specifically the $O(n^{-4})$ estimator in \citep{yakowitz1978weighted}, can be exploited in this context. We describe the Yakowitz estimator in more details in section \ref{sec:yakowitz}, and present the main idea of merging Yakowitz with Nested Sampling idea in Section \ref{sec:qis}. We take up Vertical likelihood estimation and its connection to nested sampling and posterior Lorenz curve in section \ref{sec:vertical}. Then, we show in section \ref{sec:simulation} that this estimator converges faster compared to the nested sampling estimator with $\e^{-i/N}$ weights, and na\"ive Monte Carlo Integration. We end with pointers for future work in section \ref{sec:discussion}. 

\section{Yakowitz's Riemann Sum Estimator}\label{sec:yakowitz}

\subsection{Riemann vs. Rectangular Sum}
We shall quickly review the Riemann sum estimators for evidence approximation now, and illustrate with two numerical examples. Recall that the usual method for estimating $Z$ is the na\"ive Monte Carlo, which generates random samples $x^{(i)}$ from the distribution $F$ and approximates $Z$ as $\hat{Z} = \frac{1}{n} \sum_{i=1}^n L(x^{(i)})$. The variance of $\hat{Z}$ converges at a rate of $O(n^{-1})$. The distribution $P$ might not be easy to sample from, and an alternative method to work around that is importance sampling (IS). IS involves drawing $x^{(i)}$ from a proposal distribution $g(x)$ and weighting the likelihood evaluations with $q^{(i)} \propto f(x^{(i)}) / g(x^{(i)})$. The approximation of $Z$ becomes $\hat{Z} = n^{-1}\sum_{i=1}^n q^{(i)} L(x^{(i)})$. 

Motivated by the desire to improve the convergence rate by allowing for more complicated operations on the random pairs $( x^{(i)} , L( x^{(i)} ))_{i=1}^n$, \citet{yakowitz1978weighted} proposed the use of a weighted Monte-Carlo scheme. When the prior is uniform $f = \mathbb{I}_{ [0,1] } $, they show that a convergence rate of $O(n^{-4}) $ is available for a Riemann sum estimator of the form:
\[
\hat{Z} = \sum_{i=1}^{n-1} ( u_{(i+1)} - u_{(i)} ) \frac{ L(u_{(i+1)} ) + L( u_{(i)}) }{2}
\]
where $\{u_{(i)} \}_{i = 1}^{n}$ are ordered uniform draws from $f = \mathbb{I}_{[0,1]}$. The $O(n^{-4})$ error rate follows from the fact that the joint distribution between the gaps of the uniform order statistics: $(u_{(1)}, u_{(2)}-u_{(1)}, \ldots, 1-u_{(n)})$ follows a standard Dirichlet distribution, and utilizing properties of Dirichlet moments and error bounds for trapezoidal integration rules. %We have reproduced the original proof in Appendix.  

Yakowitz's estimator \eqref{eq:yakowitz} was extended by \citet{philippe1997processing, philippe2001riemann} to the case of a general density $f(\cdot)$, with the corresponding distribution $F(\cdot)$, by taking ordered samples:
$$
x_{(1)} \doteq F^{-}(u_{(1)}) \le x_{(2)} \doteq F^{-}(u_{(2)}) \le \cdots \le x_{(n)} \doteq F^{-}(u_{(n)}),$$ and considering the Riemann sum:
\begin{align}
    \hat{\psi}_{R} & = \sum_{i=0}^{n-1} l(x_{(i)}) f(x_{(i)}) (x_{(i+1)} - x_{(i)}) \label{eq:riemann}. \\
    \text{or, equivalently} \; \hat{\psi}_{R} & = \frac{ \sum_{i=0}^{n-1} l(x_{(i)}) f(x_{(i)}) (x_{(i+1)} - x_{(i)})}{ \sum_{i=0}^{n-1} f(x_{(i)}) (x_{(i+1)} - x_{(i)})} \; \nonumber
\end{align}

The second form is useful if $f(x)$ is available only up to a multiplicative constant, which is often the case in Bayesian analysis with non-conjugate prior-posteriors. The general Riemann sum estimator above attains a convergence rate of $O(n^{-2})$, while admitting a vanishing bias of $O(n^{-1})$ order, and as \citet{philippe1997processing, philippe2001riemann} argue this convergence is `far from formal' and noticeable in cumulative sum graphs within a short range. 

Despite the much faster convergence rates for the Riemann sum estimators, particularly the Yakowitz estimator, they did not gain much popularity in evidence approximation literature as extending these trapezoid rules to higher dimension would be thwarted by the `curse of dimensionality'. An overlooked fact was that high-dimensional integrals can be written as unidimensional integrals over the positive real line \eqref{eqn:Zin1D}, or over the unit interval \eqref{eqn:Zin1D2}, as in the NS approach, the latter being the perfect situation to invoke the Yakowitz estimator. In Section \ref{sec:qis}, we show how to apply Yakowitz idea for nested sampling and argue that this estimator is worth taking another look for evidence estimation.

\subsection{Two numerical examples}
We show the faster convergence of the Riemann sum estimator over the na\"ive MC using a simulation experiment for evaluating two integrals (1) ${\rm Beta}(a,b)$ integral, for $a = 3$, $b = 3$ and (2) $\E(1/(1+U))$ for $U \sim {\rm Exp}(1)$. The true value of the for the first integral is the Beta density normalizing constant $Z = \int_0^1 x^2 (1-x)^2 \d x = {\rm Beta}(3,3)$. For the second integral the true value involves the exponential integral $E_1(x) = \int_x^{\infty} e^{-t}/t dt$, and is given by:
\[
Z = \int_0^{\infty} \frac{e^{-x}}{1+x}dx = e \cdot \int_1^{\infty} \frac{e^{-t}}{t}dt = e \cdot E_1(1). 
\]
The na\"ive Monte Carlo evaluate this integral by generating uniform samples and evaluating the Beta density function for each random sample and taking the empirical average. The Yakowitz Riemann sum \eqref{eq:yakowitz} estimates the same by first ordering the uniform samples and then applying the trapezoidal rule over them. For the second integral, we apply the Riemann sum estimator \eqref{eq:riemann} proposed by \citet{philippe1997processing}. We sample from the standard exponential distribution, order the obtained sample and apply the trapezoid or the Riemann sum formula. 

\begin{figure}[H]
    \centering
\begin{subfigure}[t]{0.49\textwidth}
    \includegraphics[width= \textwidth]{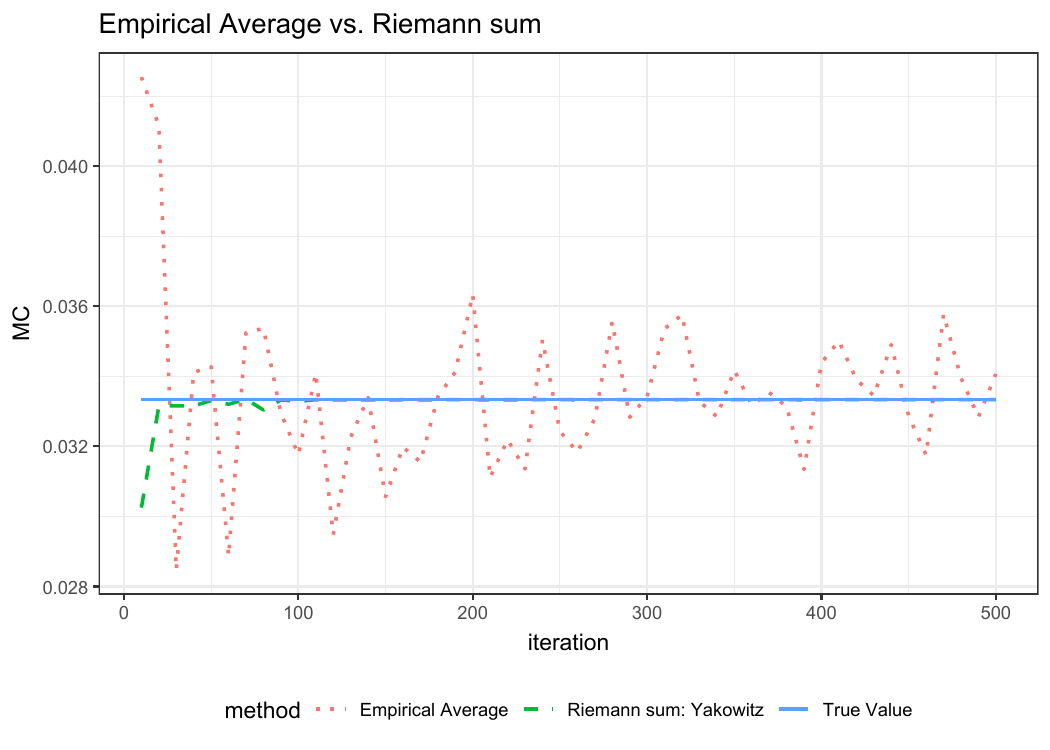}
    \caption{Estimating ${\rm Beta}(3,3)$}
    \label{fig:riemann1}
\end{subfigure}
\begin{subfigure}[t]{0.49\textwidth}
    \includegraphics[width= \textwidth]{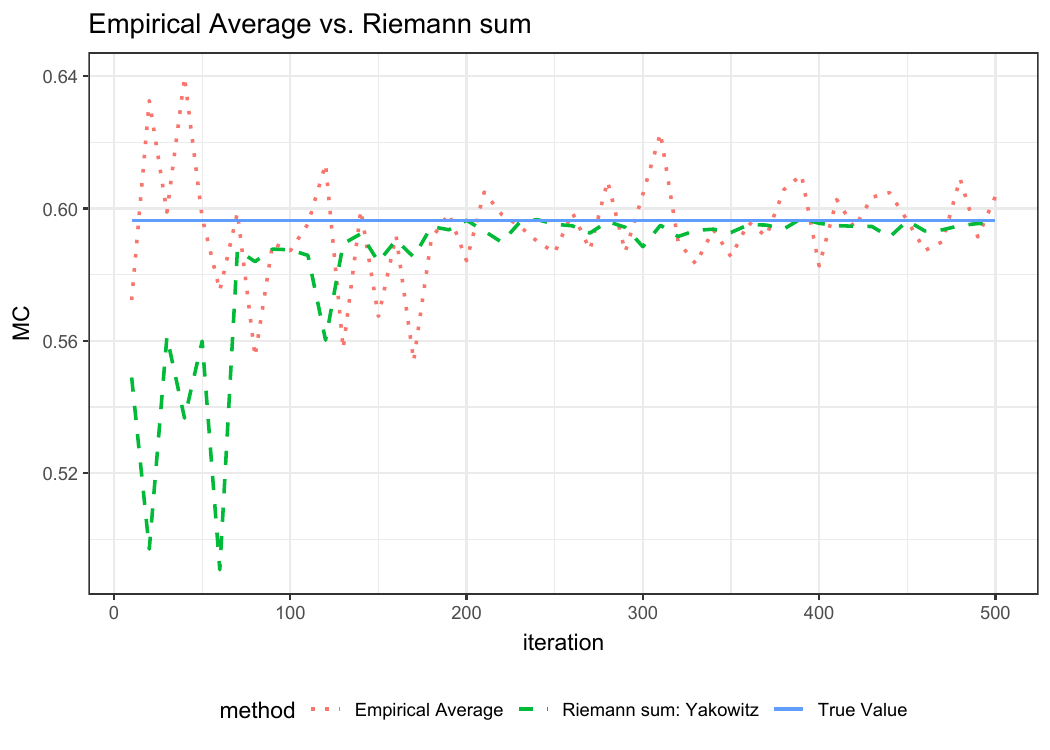}
    \caption{Estimating $\E(1/(1+U))$ for $U \sim {\rm Exp}(1)$}
    \label{fig:riemann2}
\end{subfigure}
    \caption{Convergence of estimators as a function of number of iterations for estimating integrals by various methods: Riemann sum (Yakowitz), Na\"ive Monte Carlo (empirical average), and the true value.}
    \label{fig:riemann}
\end{figure}

The result can be seen in Fig. \ref{fig:riemann}, with the Beta integral example in Fig. \ref{fig:riemann1} and the Exponential integral in Fig. \ref{fig:riemann2}. Note that the Yakowitz integral converges to the truth within a much smaller number of simulated draws due to its $O(n^{-4})$ rate, and the na\"ive Monte Carlo the slowest $O(n^{-1})$ rate, and the general Riemann sum estimator in the second example converges faster at a $O(n^{-2})$ rate. We point the readers to \citet{philippe1997processing, philippe2001riemann} for more illuminating examples and a proof. 

\section{Quantile Importance Sampling}\label{sec:qis}

\subsection{Quantile Reordering Trick} 
We start with the Lorenz identity, stated as a lemma, that is at the heart of nested sampling and the quantile importance sampling. 
\begin{lemma}\label{lemma:lorenz}
    Since $\X \sim F$, and $U \sim \UnifRV(0,1)$, it follows that $\Lambda(U) \iidd L(\X)$, that is the likelihood ordinates are distributionally same as the $\Lambda(u)$ values at uniform grid points. 
    It also follows from the definition of $\Lambda \equiv Z^{-1}$, that $Z\{L(\X)\} \sim \UnifRV(0,1)$.
\end{lemma}

Let us consider the slightly modified alternative presentation of nested sampling by \citet{polson2014vertical}. In \citep{polson2014vertical}, the  evidence $Z = \int_0^1 \Lambda(s) \d s$ is approximated at $n_{\rm iter} \doteq n$ grid points $1 \equiv s_0 > s_1 > s_2 > \cdots > s_{n} > 0)$ by a numerical integration rule: 
\begin{align}
\hat{Z}_{NS} & = \half \sum_{i=1}^{n}w_i \Lambda(s_i) \; \text{for} \; w_i = (s_{i-1} - s_{i}) \; (\text{simple}) \; \text{or} \; w_i = \half(s_{i-1} - s_{i+1}). \; (\text{trapezoid}) \label{eq:ns-polson} 
  %\\
%= \half \sum_{i=1}^{n_{\rm iter}}w_i \{\Lambda(s_{i-1})+\Lambda(s_i)\} \; \text{for} \; w_i = s_{i-1} - s_{i}. %\; \text{or} \; w_i = \half(s_{i-1} - s_{i+1}). \label{eq:ns-skilling}
% 
\end{align}

The trapezoidal rule can also be written as $\widehat{Z}_{\mathrm{NS}}
= \tfrac{1}{2}\sum_{i=1}^{n} (s_{i-1}-s_i)\{\Lambda(s_{i-1})+\Lambda(s_i)\},$ with the deterministic grid $s_i=\exp(-i/n)$ for $i=1,\ldots,n$.
For nested sampling we typically take a large number of grid points $n$ and a modest number of live points $m$;
\citet{polson2014vertical} recommend $n=1000$ and $m=20$. For a generic problem, \citet{ashton2022nested} recommends choosing the number of live points bearing in mind the trade-off between increased run-time and decreasing uncertainty (or increasing precision), and the possible multi-modal nature of the constrained prior.

The rationale for the deterministic choice $s_i = \exp(-i/n)$ follows from two facts: (a) the `volume' variable $Z(y)$ is uniformly distributed (see Lemma \ref{lemma:lorenz}) and (b) the $i^{th}$ order statistics from $n$ $\UnifRV(0,1)$ random sample has a $\BetaRV(i, n-i+1)$ distribution. It follows that, at every step, the outermost sample discarded will be distributionally same as the maximum order statistics of $n$ IID $\UnifRV(0,1)$ random variables, distributed as a $\rm{Beta}(n,1)$ random variate. This leads to $\E(s_i) = n/(1+n) \approx \e^{-1/n}$. Since the subsequent grid points can be thought of as maximal order statistics of $n$ draws from a scaled uniform: $s_0 \equiv 1$, $s_i = t_i s_{i-1}$, where $t_i$'s are the compression factors at each step and $\E(\log(t_i)) \approx -1/n$, leading \citet{skilling2006nested}'s `crude implementation' with $s_i = \exp(-i/n)$. The \textit{usual} approximation for evidence $Z$ is now straightforward, using any numerical integration rule like \eqref{eq:ns-polson}, using a rectangular or trapezoidal rule. It is important to remember that the knowledge of the functional form of $\Lambda(\cdot)$ is \textit{not required} as one can simulate points on the Lorenz curve iteratively, as long as one can simulate from the prior distribution subject to a constraint on the likelihood. We discuss this in the next section. 

Now, let us consider the Yakowitz estimator of the evidence. Let $\{ U_{(i)} \}_{i=1}^n$ denote $n$ ordered draws from $\UnifRV(0,1)$, with augmented fixed boundary points: $U_{(0)} \equiv 0$ and $U_{(n+1)} \equiv 1$. We can think of these as stochastic grid points on $[0,1]$, with $s_i = U_{(n+1-i)}$, with $s_{n+1} \equiv U_{(0)}$, and $s_0 = U_{(n+1)}$. Then the Yakowitz estimator, or the Quantile Importance Sampling estimator, will be:
\begin{gather}
    \hat{Z}_{QIS} = \sum_{i=1}^{n} w_i \Lambda(u_{(i)}) \text{with} \; w_i = U_{(i)}- U_{(i-1)}, \; (\text{simple}) \; \text{or} \; w_i = \half(U_{(i+1)} - U_{(i-1})) \; (\text{trapezoid}). \label{eq:ns-yakowitz}
    % \hat{Z}_{QIS} = \half \left[ (1-s_1) \Lambda(1) + \sum_{i=1}^{n} (\Lambda(s_{i}) + \Lambda(s_{i+1}))(s_i - s_{i+1}) + s_n \Lambda(s_n) \right] \; \text{with} \; s_i = U_{(n+1-i)}. \label{eq:ns-yakowitz}
    % \hat{Z}_{QIS} = \half \left[ (1-U_{(n)} \Lambda(U_{(n)}) + \sum_{i=1}^{n}\{\Lambda(U_{(n-i)}) + \Lambda(U_{(n-i+1)})\}\{ U_{(n-i+1)} - U_{(n-i)}\} + (1- U_{(n)}) \Lambda(U_{(n+1)}) \right].
\end{gather}

% Note that, both $Z$ and $Z_{QIS}$ admits natural upper and lower bounds: 
% \begin{equation}
%     \sum_{i=1}^{n}  (U_{(i)}- U_{(i-1)}) \Lambda(U_{(i)}) \le Z_{QIS} \le \sum_{i=1}^{n}  (U_{(i+1)}- U_{(i)}) \Lambda(U_{(i)}) + U_{(1)} \Lambda_{\rm max}. \label{eq:ns-bound}
% \end{equation}

Now, from Lemma \ref{lemma:lorenz}, we know that $\Lambda(U) \iidd L(\X)$, that is we can replace likelihood ordinates with $\Lambda(u)$ values at uniform grid points\footnote{From the corollary: $\Lambda \equiv Z^{-1}$, it follows that $Z\{L(\X)\} \sim \UnifRV(0,1)$, which leads to the correspondence of Vertical Likelihood and Nested Sampling, as discussed in \citet{polson2014vertical}.}. Put in our context, to apply Yakowitz's method for nested sampling, the main trick is to \textbf{reorder the quantiles of the likelihood ordinates}. In other words, all we need are: 
\begin{enumerate}
    \item Draw $m$ samples from the prior $\x_1, \ldots, \x_{m} \sim p(\x)$, 
    \item Calculate the likelihood ordinates at the prior values: $Y_1 = L(x_1), \ldots, Y_{m} = L(x_{m})$, and sort the sequence of $Y_j$'s obtained thus, \textit{i.e.} $Y_{(1)}, \ldots, Y_{(m)}$, 
    \item Generate $n (< m)$ grid points from $\UnifRV(0,1)$, and sort them: $u_{(1)} < u_{(2)} < \ldots < u_{(n)}$. Augment $u_{(i)}$'s with $u_{(0)} = 0$, and $u_{(n+1)} = 1$. Evaluate the sample quantiles for $Y$ at the points $u_{(i)}$: $\Lambda(u_{(i)}) \iidd Y_{\lceil m u_{(i)} \rceil}$, $i = 1, \ldots, n$, with $\Lambda(0) = \Lambda_{\rm max}$ and $\Lambda(1) = \Lambda_{\rm min}$, and use the QIS estimator given by formula \eqref{eq:ns-yakowitz}. 
% \begin{multline}
\begin{equation}
\hat{Z}_{QIS} = \left[\sum_{i=1}^{n+1} (u_{(i)} - u_{(i-1)}) \frac{\Lambda(u_{(i-1)}) + \Lambda(u_{(i)})}{2} \right].\label{eq:ns-yakowitz-2}
% \hat{Z}_{QIS} = \left[u_{(1)} \frac{(\Lambda(u_{(1)}) + Y_{\rm min})}{2} + \sum_{i=1}^{n} (u_{(n-i+1)} - u_{(n-i)}) \frac{\Lambda(u_{(n-i)}) + \Lambda(u_{(n-i+1)})}{2} + (1- u_{(n)}) \frac{Y_{\rm max}}{2} \right].\label{eq:ns-yakowitz-2}
\end{equation}
% \end{multline}

% \begin{equation}
%     \hat{Z}_{Y} = \half \left[ U_{(1)} Y_{(\lceil 0 \rceil)}+ \sum_{i=1}^{n} Y_{(\lceil mU_{(i)} \rceil)}\{ U_{(i+1)} - U_{(i)}\} + (1- U_{(n)}) Y_{(\lceil m \rceil)} \right]. \label{eq:ns-yakowitz-2} 
%     % \hat{Z}_{Y} = \half \left[ U_{(1)} \Lambda(U_{(0)}) + \sum_{i=1}^{n} \Lambda(U_{(i)}) \{ U_{(i+1)} - U_{(i)}\} + (1- U_{(n)}) \Lambda(U_{(n)}) \right]. \label{eq:ns-yakowitz-2} 
% \end{equation}
\end{enumerate}
To emphasize the critical role of quantiles in evaluating the evidence, we call this `Quantile Importance Sampling' (abbreviated as QIS). 

% \newpage

\subsection{Convergence Rate for Quantile Importance Sampling}

We show here that the quantile importance sampling estimator in \eqref{eq:ns-yakowitz} can achieve a $O(n^{-4})$ rate of convergence for evidence using the results of \citep{yakowitz1978weighted}. We restate the results here using the nested sampling terminology. Let $\Lambda(s)$ be the pseudo-inverse of the upper cumulant of the likelihood ordinate, in the Yakowitz formula for nested sampling \eqref{eq:ns-yakowitz}: then the Yakowitz lemma for nested sampling can be stated as follows: 
\begin{lemma}\label{prop:1}
Let the evidence $Z \equiv \int_{\chi} L(\x) dF(\x) = \int_0^1 \Lambda(s) ds$. Assume that the cumulant $\Lambda(s)$ has a continuous derivative and that the second derivative $\Lambda''(s)$ exists and is bounded in absolute value over the unit interval. Let $U_{(0)} \equiv 0, U_{(n+1)} \equiv 1$, and $\{U_{(i)}\}_{i=1}^n$ be the ordered statistics from $\UnifRV(0,1)$ distribution (with $U_{(i)} \ge U_{(i-1)}, i = 1, \ldots, n+1$). Consider the trapezoid estimator defined in \eqref{eq:ns-yakowitz}:
\[
\hat{Z}_{QIS} = \half \left(\sum_{i=1}^{n+1} (U_{(i)} - U_{(i-1)}) \{ \Lambda(U_{(i-1)}) + \Lambda(U_{(i)}) \} \right).
\]
% \begin{align*}
%     \hat{Z}_{n} & = \frac{1}{2} \left[ \sum_{i=1}^{n+1} \left\{ \Lambda(U_{(i-1)}) + \Lambda(U_{(i)}) \right\} (U_{(i)} - U_{(i-1)})\right] \le \tilde{Z}_{n} \\
% \end{align*}
The following hold true: 
\begin{enumerate}
    \item Firstly, for some constant $M$,
\[
\E[(Z - \hat{Z}_{QIS})^2] \leq M/n^4, \qquad \text{for all } n \geq 1. 
\]
\item $Z_{QIS}$ admits natural upper and lower bounds: 
\begin{equation}
    \sum_{i=1}^{n}  (U_{(i)}- U_{(i-1)}) \Lambda(U_{(i)}) \le \hat{Z}_{QIS} \le \sum_{i=1}^{n}  (U_{(i+1)}- U_{(i)}) \Lambda(U_{(i)}) + U_{(1)} \Lambda_{\rm max}. \label{eq:ns-bound}
\end{equation}
\end{enumerate}
\end{lemma}
We make a couple of remarks about the rate before presenting the result. Note that, $\Lambda(u)$ is non-increasing in $u$ and $\Lambda(0)$ corresponds to the maximum value of the likelihood ordinate, denoted as $\Lambda_{\rm max}$ in \eqref{eq:ns-bound}. 

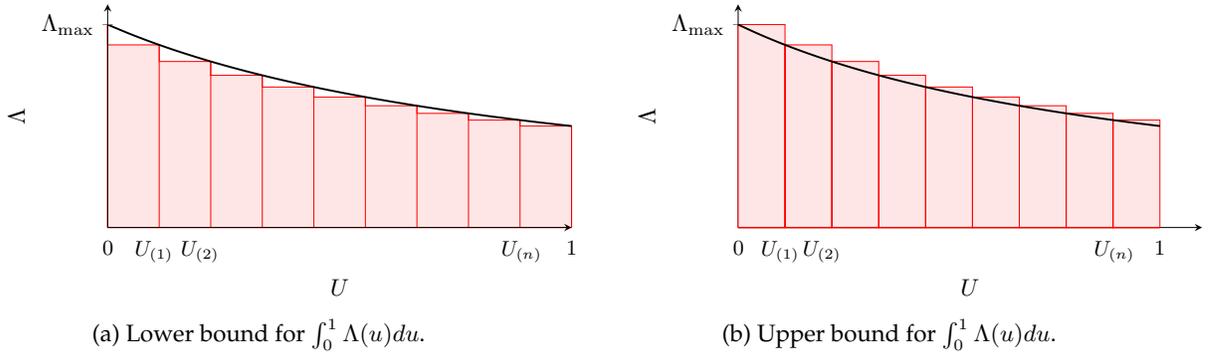
\begin{figure}[h!]
    \centering
    \begin{subfigure}[t]{0.45\linewidth}
    \centering
        \begin{tikzpicture}[scale=0.9]
        \begin{axis}[
        y=3cm, xmax=1, ymax=1.1,ymin=0,xmin=0,
        xtick={0.001, 0.1, ..., 0.9, 1}, ytick={0,1},
        xticklabels={$0$, $U_{(1)}$, $U_{(2)}$,  , ,,,,,$U_{(n)}$, $1$},
        yticklabels={$0$, $\Lambda_{\rm max}$},
        x tick label style={font=\small, rotate=0, below},
        enlargelimits=true,
        axis lines=middle,
        clip=false,
        domain=0:1,
        axis on top,
        xlabel={$U$},
        ylabel={$\Lambda$},
        ylabel style={rotate=-90},
            ylabel near ticks,
            xlabel near ticks,
                xtick style={draw=none}
        ]
        \addplot [draw=red,fill=red!10,const plot mark right, samples=10, domain=0:1]
        {1/(1+x)}\closedcycle;
        % \addplot [draw=red, fill=red!10, ybar interval, domain = 0:0.9, samples=9]
        % {1/(1+x)}\closedcycle;
        % The vertical lines between the segments
        \addplot [draw=red, ycomb,samples=10, domain=0:1] {1/(1+x)};
        \addplot[smooth, thick,domain=0:1]{1/(1+x)};
        \end{axis}
        \end{tikzpicture}
    \caption{Lower bound for $\int_0^1 \Lambda(u) du$.}
        \label{fig:bounds-lower}
    \end{subfigure}
    \hfill
\begin{subfigure}[t]{0.45\linewidth}
\centering
    \begin{tikzpicture}[scale=0.9]
        \begin{axis}[
        xtick={0.001, 0.1, ..., 0.9, 1}, ytick={0,1},
        xticklabels={$0$, $U_{(1)}$, $U_{(2)}$,  , ,,,,,$U_{(n)}$, $1$},
        yticklabels={$0$, $\Lambda_{\rm max}$},
        x tick label style={font=\small, rotate=0, below},
        y=3cm, xmax=1.1,ymax=1.1,ymin=0,xmin=0,
        enlargelimits=true,
        axis lines=middle,
        clip=false,
        domain=0:1,
        xlabel={$U$},
        ylabel={$\Lambda$},
        ylabel style={rotate=-90},
            ylabel near ticks,
            xlabel near ticks,
                xtick style={draw=none}
        ]
        % \addplot [draw=red,fill=green!90,const plot mark right, samples=5, domain=0:0.8]
        % {1/(1+x)}\closedcycle;
        \addplot [draw=red, fill=red!10, ybar interval, domain = 0:1, samples=10]
        {1/(1+x)}\closedcycle;
        \addplot[smooth, thick,domain=0:1]{1/(1+x)};
        \end{axis}
        \end{tikzpicture}
    \caption{Upper bound for $\int_0^1 \Lambda(u) du$.}
        \label{fig:bounds-upper}
\end{subfigure}
    \caption{Lower and upper bounds for the QIS (Riemann sum estimator) of $\int_0^1 \Lambda(u) du$.}
    \label{fig:bounds}
\end{figure}

\begin{remark}\label{remark:max}
    As pointed out by \citet{skilling2006nested}, the $\Lambda_{\rm max}$ is the maximum likelihood value, which is independent of the nested sampling algorithm, and there could be intervals of negligible size containing very large likelihood values (see Fig. \ref{fig:bounds}) unless precluded by separate global analysis, and the error for numerical integration could be at most $O(n^{-1})$. 
\end{remark}

\begin{remark}\label{remark:better}
    The $O(n^{-4})$ rate of convergence using Yakowitz's lemma is a major improvement on the rate obtained by the Central Limit Theorem in \citet{chopin2010properties}. The CLT for nested sampling assumes that the $\Lambda(\cdot)$ function is twice continuously differentiable over $[\epsilon, 1]$ and its first two derivatives are bounded over $[\epsilon, 1]$, where $\epsilon > 0$ depends on the desired accuracy. On the other hand, QIS rate assumes that $\Lambda$ has a continuous derivative with $\Lambda''$ bounded over $[0,1]$. 
\end{remark}

We present the proof of Lemma \ref{prop:1} here for the sake of completeness. 

\begin{proof}
First, we use the following lemma from \citet{yakowitz1978weighted}.  
\begin{lemma}\label{lemma:1}
Let $\{U_{i}\}_{i=1}^n \sim \UnifRV(0,1)$ be a random sample from uniform distribution and let the gaps between the ordered statistics be denoted by $Z_1 = U_{(1)}$, $Z_j = U_{(j)}-U_{(j-1)},\; 2 \leq j \leq n,$ and $Z_{(n+1)} = 1-U_{(n)}$, implying that jointly $(Z_1, \ldots, Z_{(n+1)}) \sim \rm{Dir}(1, 1, \ldots, 1)$. Then, it follows that:
\begin{equation}
    \E[Z_i^6] = 6!n!/(n+6)!, \qquad \E[Z_i^3Z_j^3] = (3!)^2n!/(n+6)!, \qquad i, j = 1, 2 , \ldots, n+1,\qquad i\neq j. \label{eq:dir-moments}
\end{equation}
\end{lemma}
The proof follows from a similar argument as in \citet{yakowitz1978weighted}. It uses the well-known error bound for the Trapezoid rule for integral approximation: 
\beq
\int_a^b f(t)dt - \frac{b-a}{2} [f(b) + f(a)] = -\frac{(b-a)^3}{12} f''(\xi), \; 0 \leq a \le \xi \le b \leq 1. \label{eq:trap}
\eeq
Noting that $Z = \int_0^1 \Lambda(u) du$, and $U_{(i)}$'s are the ordered $\UnifRV(0,1)$ statistics along with $U_{(0)} \equiv 0, U_{(n+1)} \equiv 1$, we can write:
$$
Z - \hat{Z}_{QIS} = \sum_{i=1}^{n+1} \left( \int_{U_{(i-1)}}^{U_{(i)}} \Lambda(t)dt - \frac{1}{2} [\{\Lambda(U_{(i-1)}) + \Lambda(U_{(i)})\}(U_{(i)} - U_{(i-1)})] \right).
$$
where $a \leq \xi \leq b$. Using this result, we have that if $C \geq |\Lambda''(u)|,$ $0 \leq u  \leq 1$, then
\begin{eqnarray*}
\abs{Z - \hat{Z}_{QIS}} &\leq& \left| \int_0^{U_{(1)}} \Lambda(t)dt - \frac{1}{2} U_{(1)} \{ \Lambda(0) + \Lambda(U_{(1)} \}  \right|\\
& & +\left| \sum_{i=1}^{n-1} \int_{U_{(i)}}^{U_{(i+1)}} \Lambda(t)dt -  \frac{(U_{(i+1)} - U_{(i)})}{2} \{ \Lambda(U_{(i)}) + \Lambda(U_{(i+1)})\} \right|\\
& & +\left| \int_{U_{(n)}}^1 \Lambda(t) dt -  \frac{(1 - U_{(n)})}{2} \{ \Lambda(U_{(n)}) + \Lambda(1) \}  \right|\\
&\leq& \frac{C}{12} \sum_{i=1}^{n+1} Z_i^3, \qquad (\text{applying} \; \eqref{eq:trap} \; \text{repeatedly}).
\end{eqnarray*}
where the $Z_i$'s follow a symmetric ${\rm Dirichlet}(\1)$ as per Lemma \ref{lemma:1}, and we have: 
\begin{eqnarray*} 
\E[(\hat{Z}_{QIS} - Z)^2] &\leq& \E\left[ \left( \frac{C}{12} \sum_i Z_i^3 \right)^2 \right] = \frac{C^2}{144} \left[ \sum_i \E[Z_i^6] + \sum_{i \neq j} \E[Z_i^3Z_j^3] \right]\\
&=& \frac{C^2}{144} ((n + 1)n(3!)^2n!/(n +6)! +(n + 1)6!n!/(n +6)!)\\
&=& \frac{C^2}{4} \frac{n+20}{(n+2)(n+3)(n+4)(n+5)(n+6)} = O(n^{-4}).
\end{eqnarray*}
\qedhere
\end{proof}

% \citet{philippe1997processing} generalizes this to arbitrary priors $P(dx) = p(x) d x $ using an importance sampling approach and provides a convergence rate of $O(N^{-2}) $ for an estimator of the form: 
% \[
% \hat{Z} =  \sum_{i=1}^{n-1} ( x_{ (i+1) } - x_{ (i) } ) p( x_{(i)} ) L( x_{(i)} ) 
% \]
% where $ x_{(1)} \leq \ldots \leq x_{(n)} $ are the ordered sample associated with $(x_1, \ldots , x_n ) \sim P$.

\section{Vertical Likelihood and Nested Sampling}\label{sec:vertical}

\subsection{Simulating the Posterior Lorenz Curve}\label{sec:lorenz} 
% \subsection{Simulating the Lorenz Curve}\label{sec:simlorenz}

Recall that the Lorenz curve, $\mathcal{L}$ of $Y$ is defined in terms of its CDF, $F_Y(y)$, as: 
\begin{gather*}
  \mathcal{L}(u) = \frac{1}{Z} \int_0^u F_Y^{-1} ( s ) d s = \frac{1}{Z} \int_0^u \Lambda( s ) d s = \frac{1}{Z} L(u)  \; \; \text{ where } \; u \in [0,1], \\
  \text{where} \; Z = \mathbb{E}( Y) = \int_{ \mathcal{X} } L(x) p(\d x) \; .
\end{gather*}
The normalization constant, $Z$, is essentially the area under the unnormalized Lorenz curve. One feature of the Lorenz curve is that, $\mathcal{L}(1) = 1$, \textit{i.e.}, it provides a way to evaluate:
\[
\mathbb{E}(Y) = \int_{\mathcal{X}} L(x)p(dx) = \int_0^1 F_Y^{-1}(s) \d s = \int_0^1 \Lambda (1-s) \d s = \int_0^1 \Lambda (s) \d s = L(1), 
\]
where $F_Y(y) = 1 - Z(y)$ and $\Lambda(s)$ via $F_Y^{-1}(s) = \Lambda (1-s)$, as before. 

An important consideration here is that we do not need to know the $\Lambda(\cdot)$ function, provided we can recursively sample from the prior space, subject to a sequence of constraints, nested within each other. That is, simulating quantiles along the Lorenz curve of the likelihood ordinate $Y \equiv L(X)$ where $X \sim p(x)$ is done by sampling from the prior $p(x)$ and sequentially from $p(x \mid L(x) > y)$ to construct the Lorenz curve. Re-weighting this also provides a (weighted) sample from the posterior $\pi(x) = L(x) p(x)/Z$, that serves the dual purpose of both inference and model comparison. With this, we can simply view the vertical likelihood samples as an empirical realization of the Lorenz curve. \cite{goldie1977convergence} provides limit theorems for the convergence of empirical Lorenz curves.

The key is to construct a grid of likelihood ordinates for a Riemann approximation of Lemma \ref{lemma:1}. Let these likelihood contour values be denoted by: $0 = L_0^{(q)} < L_1^{(q)} < \ldots < L_k^{(q)} < \infty$ for $k = 0, 1, \ldots$. A common construction in vertical likelihood sampling is to use the $(1 - q^k)$-quantiles of $L(X)$ under $p(x)$. By definition, we then have:
\[
Z( L_k ) = q^k \; \; \text{ for } \; \; k=0,1,2, \ldots. 
\]
Starting at $p(x)$ and $L_0=0$, we can construct $L_{k+1} $ from $L_k$ by drawing the set of conditional prior distributions, $p(x \mid L(x) > L_k )$ and the relation: 
\begin{align*} 
  L_k^{(q)} \defeq & (1- q^k)^{\text{th}} \; \text{quantile of $L$ under prior } p(x) \\
  =& \; (1- q)^{\text{th}} \; \text{quantile of $L$ under the constrained prior} \; p( x \mid L(x) >  L_k )  \; . 
\end{align*}

Therefore, the key insight is that if we are able to sample sequentially from the conditional prior $p(x \mid L(x) > L)$, we can recursively create a construction for estimating this set of quantiles. An advantage of this approach is that we do not need to explicitly know $Z(L_k)$ or equivalently $F_Y(L_k)$ for any $k$. \cite{polson2014vertical} provide an alternative weighted slice sampling approach that exploits the knowledge of $Z(y)$.

We also simulate $p_{\Lambda}(s)$, and calculate and approximation to $ Z= \int_0^1 \Lambda (s) ds$. A natural choice is a geometric grid  $s_k = 1-q^k $ where $q$ is given, say $q = 0.90$ for $k=1, \ldots, N$. An unbiased estimate of the quantile difference is given by:
\[
\mathbb{E} \left ( L_k^{(q)}  \right ) = \Lambda ( q^{k} ) - \Lambda ( q^{k-1} ).
\]
Using a discretization of our fundamental lemma gives a discrete approximation:
\begin{align*}
  \pi_{L^{(q)}} (x) &= \sum_{k=1}^\infty \frac{ Z(L_k) (L_k - L_{k-1} )}
    {\sum_{k=1}^N Z( L_k ) (L_k - L_{k-1} ) } p( x | L(x) > L_k) \\
                  &= \sum_{k=1}^\infty  \frac{ q^k (L_k - L_{k-1} )}
    {\sum_{k=1}^N q^k (L_k - L_{k-1} ) } p( x | L(x) > L_k)
\end{align*}
We also have the point-wise limit, $\lim_{q \to 1} \pi_L^{(q)} (x) = \pi_L(x) $ and simulated $Z^{(q)}_L \to Z$. When the quantity of interest is the normalization constant $Z$, we can obtain a simple estimate from the identity:
\beq
\hat{Z}_N = \sum^N_{k=1} q^k (L_k - L_{k-1}). \label{eq:simple}
\eeq
An asymptotic estimate is given by concatenating terms as
\beq
\hat{Z}_\infty = (1-q) \sum^\infty_{k=1} q^k L_k \;. \label{eq:asymp}
\eeq

\section{Simulation study}\label{sec:simulation}

\subsection{Normal prior, Normal likelihood} 
As an illustrative example, we take Gaussian likelihood and Gaussian prior, \textit{i.e.,} 
\begin{align*}
L(x \mid \mu_1, \sigma_1) & = \frac{1}{\sqrt{2 \pi}\sigma_1} \e^{-\frac{(x-\mu_1)^2}{2\sigma_1^2}},\\
p(x \mid \mu_2, \sigma_2) & = \frac{1}{\sqrt{2 \pi}\sigma_2} \e^{-\frac{(x-\mu_2)^2}{2\sigma_2^2}}.
\end{align*}
The integral $Z = \int L(x) p(x) dx$ is available in closed form due to the self-conjugacy property of Gaussian prior with the mean parameter of a Gaussian distribution. To see this, note that we can write the product of two Gaussian density functions in the following way: 
\begin{gather}
    \phi(x \mid \mu_1, \sigma_1) \times \phi(x \mid \mu_2, \sigma_2) = \phi(\mu_1 \mid \mu_2, \sqrt{\sigma_1^2 + \sigma_2^2}) \times \phi(x \mid \mu_{\text{post}}, \sigma_{\text{post}}^2) \\
\text{where} \quad \mu_{\text{post}} = \frac{\sigma_1^{-2} \mu_1 + \sigma_2^{-2} \mu_2}{\sigma_1^{-2} + \sigma_2^{-2}}, \sigma_{\text{post}}^2 = \frac{\sigma_1^2 \sigma_2^2}{\sigma_1^2 + \sigma_2^2}.
\end{gather}
Integrating the right-hand-side will leave the normalizing constant, here expressed as a Gaussian density itself. Let $\mu_1 = 2, \mu_2 =0$ and $\sigma_1^2 = \sigma_2^2 = 1$. The normalizing constant is:
\[
Z = \phi(2 \mid 0, 2) = \frac{1}{\sqrt{2 \pi} \sqrt{2}} \e^{-1} = \frac{1}{2 \e \sqrt{\pi}}.
\]
To implement the nested sampling algorithm, we need to calculate the survival function or upper cumulant of the likelihood ordinates, that are available as follows: 
\begin{align}
Z(y) &= \mathbb{P}\{L(X) > y\} = \mathbb{P}_{X \sim \NormRV(\mu_2, \sigma_2)}\left( \phi(X \mid \mu_1, \sigma_1) > y \right) \nonumber \\
&= \mathbb{P}\left\{ \left(\frac{X -\mu_1}{\sigma_1}\right)^2 \leq -2\log(y\sqrt{2\pi}\sigma_1) \right\} \nonumber \\
& = \mathbb{P}\left\{\mu_1 - \sigma_1 \sqrt{-2\log(y\sqrt{2\pi}\sigma_1)} \leq X  \leq \mu_1 + \sigma_1 \sqrt{-2\log(y\sqrt{2\pi}\sigma_1)} \right\} \label{eq:truncnorm} \\
& = \Phi\left(\frac{\mu_1 + \sigma_1 \sqrt{-2\log(y\sqrt{2\pi}\sigma_1)} - \mu_2}{\sigma_2} \right) - \Phi\left(\frac{\mu_1 - \sigma_1 \sqrt{-2\log(y\sqrt{2\pi}\sigma_1)} - \mu_2}{\sigma_2} \right) \nonumber
\end{align}
where $\Phi$ is the $\NormRV(0,1)$ cumulative distribution function. For the special case, $\mu_1 = 2, \mu_2 =0$ and $\sigma_1^2 = \sigma_2^2 = 1$, we get a simpler form: $Z(y) = \Phi(2 + \zeta) - \Phi(2 -\zeta)$, where $\zeta = \sqrt{-2\log(y\sqrt{2\pi})}$. 

% \begin{wrapfigure}{r}{0.5\textwidth}
%   \begin{center}
%     \includegraphics[height = 1.8in, width=0.45\textwidth]{art/lambda_100}
%   \end{center}
%   \caption{The Lambda function $\Lambda(s)$}
%   \label{fig:lambda}
% \end{wrapfigure}

\begin{figure}[ht!]
% \vskip 5pt
\centering
\includegraphics[height = 2in]{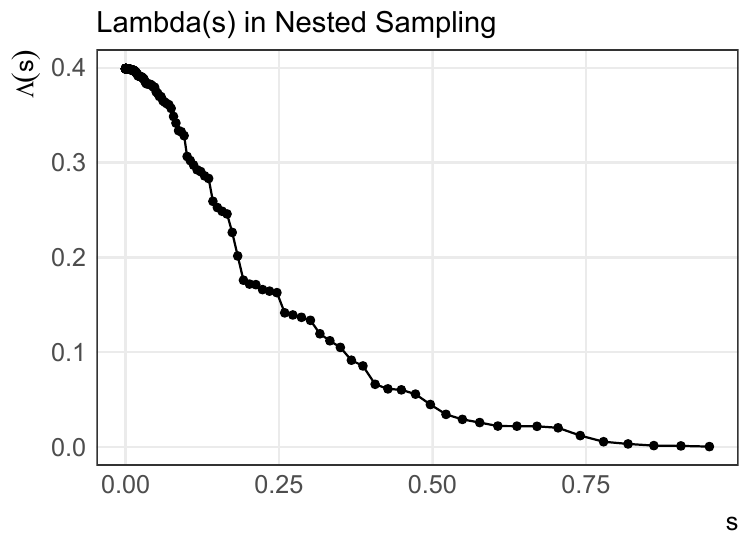} 
\caption{The Lambda function $\Lambda(s)$}
\label{fig:lambda}
\end{figure}

Clearly, the pseudo-inverse $\Lambda(s) = Z^{-1}(s)$ does not admit a closed-form expression, but as we argued earlier, it is not a hindrance in implementing the nested sampling algorithm as we can draw samples from the truncated normal distribution shown in \eqref{eq:truncnorm}. The path of the simulated $\Lambda(s)$ for the original nested sampling scheme using exponential weights is shown in Fig. \ref{fig:lambda}. One can also simulate posterior Lorenz curve along a geometric grid-points using the steps outlined in \ref{sec:lorenz}. 

% \vspace{0.1in}
% \noindent 

\begin{figure}[h!]
\centering
% \begin{subfigure}[T]{0.45\linewidth}
\includegraphics[height = 3.5in]{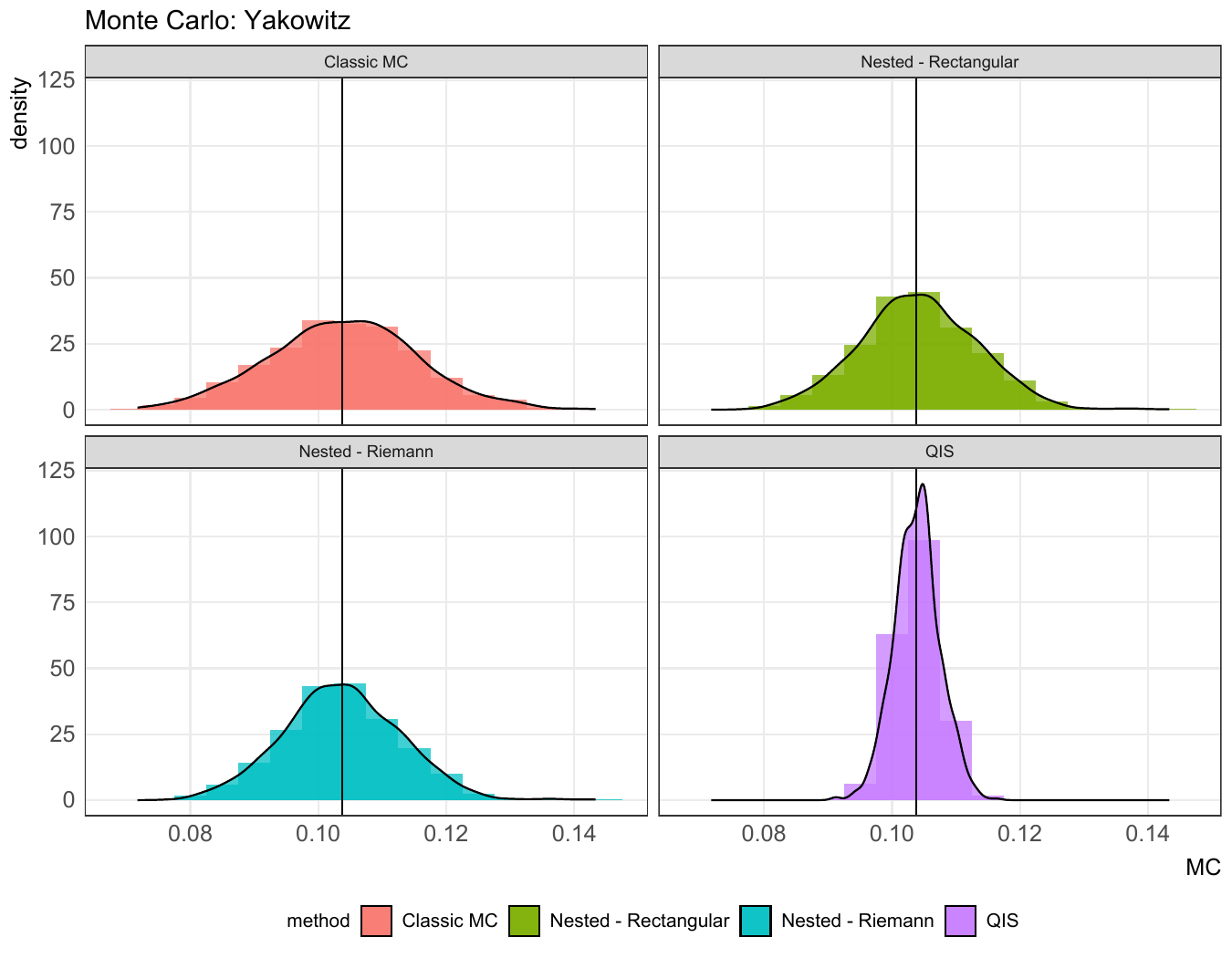} 
% \caption{}
\caption{Comparison of three candidate schemes: Yakowitz or QIS (ordered uniform), Skilling's original scheme ($\exp(-i/N)$) and the classic Monte Carlo approach.}
\label{fig:yakowitz}
% \end{subfigure}
\end{figure}

To calculate $Z$, we follow the Quantile Importance Sampling (QIS), \textit{i.e.}, the vertical likelihood idea combined with Yakowitz estimation as outlined in \ref{sec:yakowitz}, and compare with the traditional Monte Carlo approach, as well as the original nested sampling approach with both Riemann summation and rectangular summation. We take $n = 20$ live points for $m = 1000$ samples from the prior, and $100$ replicates to calculate the root mean squared errors (RMSE), $\{1/r\sum_{j=1}^{r} (\hat{Z}_j - Z)\}^{1/2}$ and mean absolute predictive error (MAPE), $1/r\sum_{j=1}^{r} \abs{(\hat{Z}_j - Z)/Z}$. Figure \ref{fig:yakowitz} shows the kernel density estimates and histograms for the $100$ $Z$-estimates obtained via the four competing approaches. Here, all methods perform relatively well with the replicates centered around the true $Z$, with different variability. Notably, the QIS approach with Riemann summation concentrate in a narrower region around the true $Z$. compared to all the other candidates, an outcome of the faster convergence to the truth. The \textsc{R} program for this experiment is available at this address: \url{https://dattahub.github.io/qis/QIS.html}. 

For the Yakowitz estimator (Quantile Importance Sampling), we sample $m$ points from the prior $X \sim \NormRV(\mu_1, \sigma_1)$, evaluate the likelihood ordinates $Y \equiv L(x) = \phi((x-\mu_2)/\sigma_2)$ at $n$ live points. Calculate the $n$ sample quantiles for these $Y$ ordinates: these are our $\Lambda(\cdot)$ values. Finally, apply the Trapezoid sum formula on the $(u, \Lambda(u))$: abscissa-ordinate pairs. For the nested sampling estimates, we repeatedly draw from the induced distribution on the likelihood ordinates given by the truncated normal in \eqref{eq:truncnorm}, calculating $\hat{Z}_{NS}$ on the grid $s_i = \exp(-i/n)$ via the rectangular and trapezoid rules in \eqref{eq:ns-polson}.

\begin{table}[h!]
    \caption{Comparison of RMSE and MAPE across three methods. True value $Z = 0.1037769$. Bold fonts indicate lowest RMSE and MAPE for this simulation.}
    \label{tab:mse_table}
    \centering
    \footnotesize{
\begin{tabular}{@{}lccccc@{}}
\hline
  & QIS & Naive & Nested-Rect & Nested-Riemann\\
\hline
Mean & 0.1039080 & 0.1038586 & 0.1037008 & 0.1042196\\
\hline
MAPE & \textbf{0.0214313} & 0.0730794 & 0.0583018 & 0.0583487\\
\hline
RMSE & \textbf{0.0035507} & 0.0115004 & 0.0090824 & 0.0091383\\
\hline
\end{tabular}
}
\end{table}

\begin{figure}[h!]
\centering
\includegraphics[height = 3.5in]{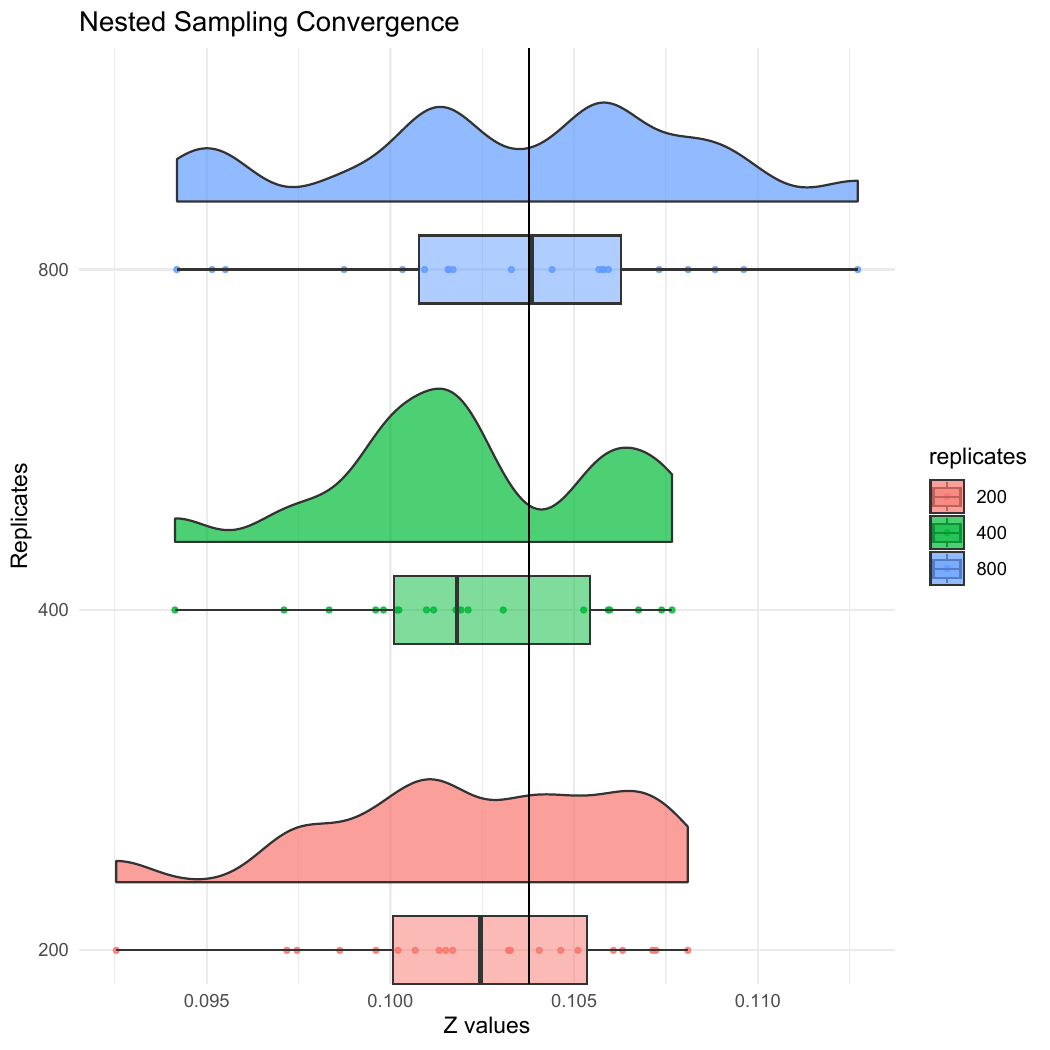} 
% \caption{}
\caption{Density and box plots (aka rain cloud plots) of vertical sampling estimates $\hat Z_N$ for $n = 200, 400, 800$ compared to true value of evidence $Z$ (solid vertical line). Individual dots indicate $20$ simulation replicates.}
\label{fig:boxplot}
% \end{subfigure}
% \caption{(a) Comparison of three candidate schemes: Yakowitz or QIS (ordered uniform), Skilling's original scheme ($\exp(-i/N)$) and the classic Monte Carlo approach, (b) Density and box plots (aka rain cloud plots) of vertical sampling estimates $\hat Z_N$ for $n = 200, 400, 800$ compared to true value of evidence $Z$ (solid vertical line). Individual dots indicate $20$ simulation replicates.}
\end{figure}

Finally, we show the effect of the number of live points on the accuracy of nested sampling. Figure \ref{fig:boxplot} shows the convergence of $\hat Z_n$ obtained via the deterministic nested sampling algorithm as a function of $n$, the number of live points. Here, we apply the vertical sampling estimator on a deterministic exponential grid where $x^{(i)} = e^{-i/m}$, $i = 0, 1,\ldots, n$, with $m = 1000$. We calculate the nested sampling estimates for $n = 200, 400, 800$ and show the estimates in side-by-side box-plots in Figure \ref{fig:boxplot}. The precision of estimates improves every time we double the number of grid points $n$, with a decreasing bias as we increase $n$. 

% \begin{subfigure}[T]{0.45\linewidth}

\subsection{Multivariate normal prior, $t$ likelihood} 
As an example of a higher-dimensional integral, we look at a multivariate $t$ likelihood and a multivariate Gaussian prior with dimension $d = 50$, following \citet{polson2014vertical}. The target integral is:
\begin{equation}
    Z = \int_{\mathbb{R}^d} (1 + \frac{\x^T \x}{\nu})^{-\frac{\nu + d}{2}} (\frac{\tau}{2\pi})^{d/2} \exp\{-\tau \x^T \x/2\} d \x. 
\end{equation}
As \citet{polson2014vertical} show, this integral can be written in terms of Kummer's confluent hypergeometric function of the second kind: $Z = s^{a}U(a, b, s)$, where $a = (\nu + d)/2$, $b = \nu/2 + 1$, $s = \nu \tau/2$. For the specific values used here: $d = 50, \tau =1$ and $\nu = 2$, we can get: $Z = U(26, 2, 1) = 1.95\times 10^{-29}$. \citet{polson2014vertical} showed that the nested sampling and weighted slice sampling both work well for this example and both the Harmonic mean estimator \citet{raftery2006estimating} and the Chib's \citep{chib1995marginal} method suffer. We complement their analysis by adding the na\"ive MC and the Yakowitz estimator or quantile importance sampling for this comparison. For this example, we take the number of live points $n = 20$, for $m = 10,000$ points and $100$ replicates. 
\begin{figure}[h!]
    \centering
    \includegraphics[height = 3in]{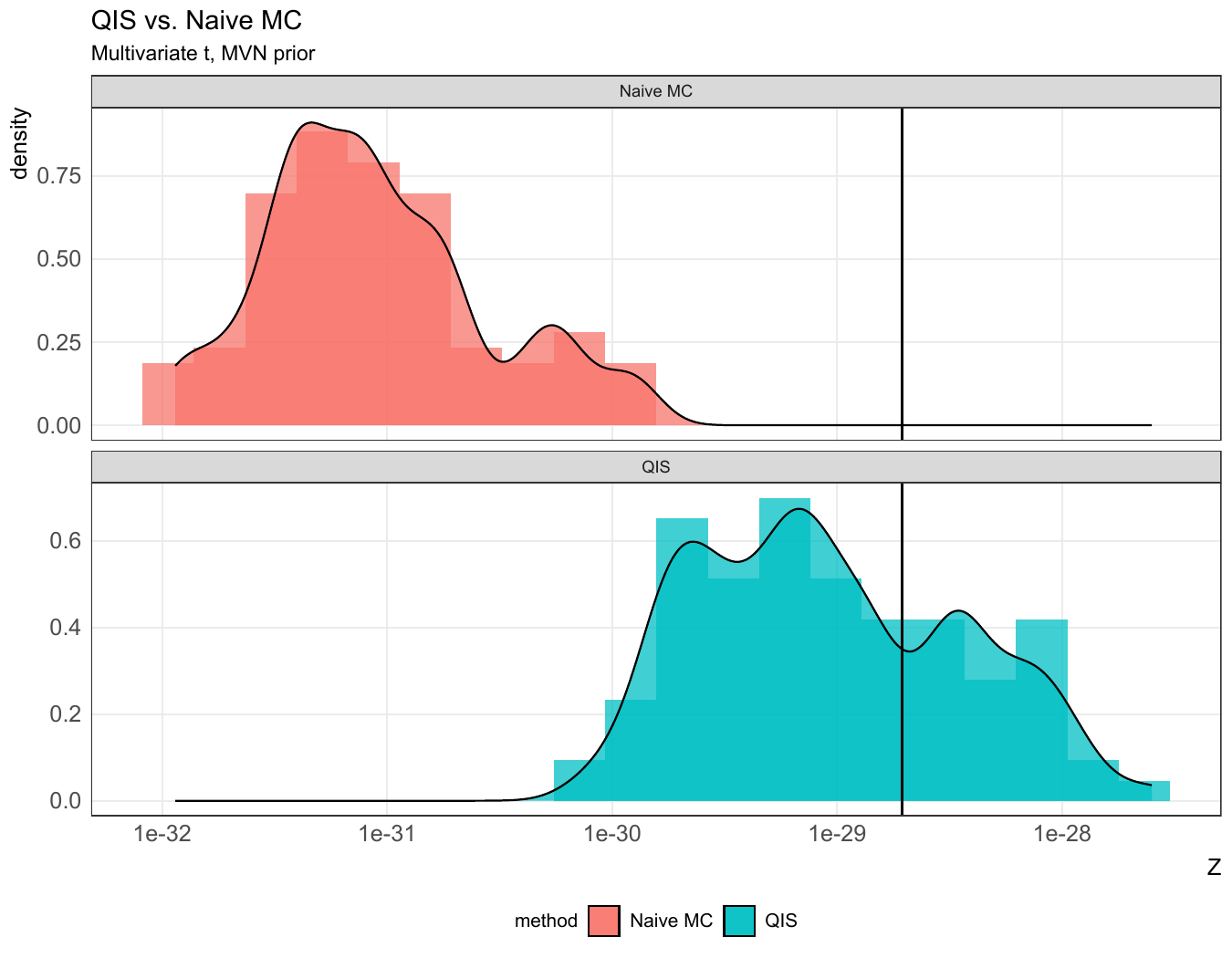}
    \caption{QIS vs. Na\"ive MC comparison for the multivariate $t$-multivariate Normal example.}
    \label{fig:ns_mvt}
\end{figure}

\begin{table}[h!]
\caption{Comparison of QIS vs. Na\"ive MC. The true value is $Z = 1.95\times 10^{-29}.$}
\label{tab:mvt}
    \centering
\footnotesize{
\begin{tabular}{@{}lcc@{}}
\hline
 Metrics & Yakowitz (QIS) & Na\"ive\\
\hline
Mean & $2.960234 \times 10^{-29}$ & $2.007210\times 10^{-31}$\\
% \hline
% Median & 7.628959e-30 & 7.598928e-32\\
\hline
MAPE & 0.8271503 & 0.9960922\\
\hline
RMSE & $4.175335\times 10^{-28}$ & $1.899170\times 10^{-29}$\\
\hline
\end{tabular}}
\end{table}
As Table \ref{tab:mvt} shows, the QIS method performs much better both in terms of the accuracy and the errors (RMSE and MAPE). The QIS estimate as well as the RMSE is in the same order as the true $Z$, indicating both high accuracy and precision. The \textsc{R} program for this experiment is available at this address: \url{https://dattahub.github.io/qis/QIS_MVT.html}. 

\subsection{Real Data: Arsenic wells data (Bangladesh)}

\noindent We consider the \texttt{wells} dataset (Bangladesh arsenic study; $n=3020$ households) analyzed in \citep{GelmanHill2007} as well as \citep{chopin2010properties}. Here, the binary response $y_k\in\{0,1\}$ records whether household $k$ switched to a safe well. The linear predictor uses seven covariates as in Skilling’s benchmark \citep{SkillingTests}: $x_1$ distance to the nearest safe well in 100 m units, $x_2=\log(\text{arsenic})$ in mg/L, $x_3 = $ education evel from $0$ to $17$ divided by 4, relative to the mean, the interactions $x_4=x_1x_2$, $x_5=x_1x_3$, $x_6=x_2x_3$, and an intercept $x_7\equiv1$. With a probit link $s_k=\sum_{j=1}^{7}x_{kj}\theta_j$, the likelihood is $L(\theta)=\prod_{k=1}^{n}\{\Phi(s_k)\}^{y_k}\{1-\Phi(s_k)\}^{1-y_k}$ and, for the baseline, the prior is $\theta_j\sim N(0,10^2)$ independently. \citet{SkillingTests} reports a stabilized level $\log Z=-1969.552$ with information $H=34.208$ and, over 100 replicated nested-sampling runs, mean (sd) values for $\log Z$ of $-1970.40\,(6.12)$ for $N=1$ live point, $-1969.72\,(1.64)$ for $N=10$, $-1969.57\,(0.63)$ for $N=100$, and $-1969.55\,(0.18)$ for $N=1000$. 
\begin{figure}[h!]
\centering
\includegraphics[width=.5\linewidth]{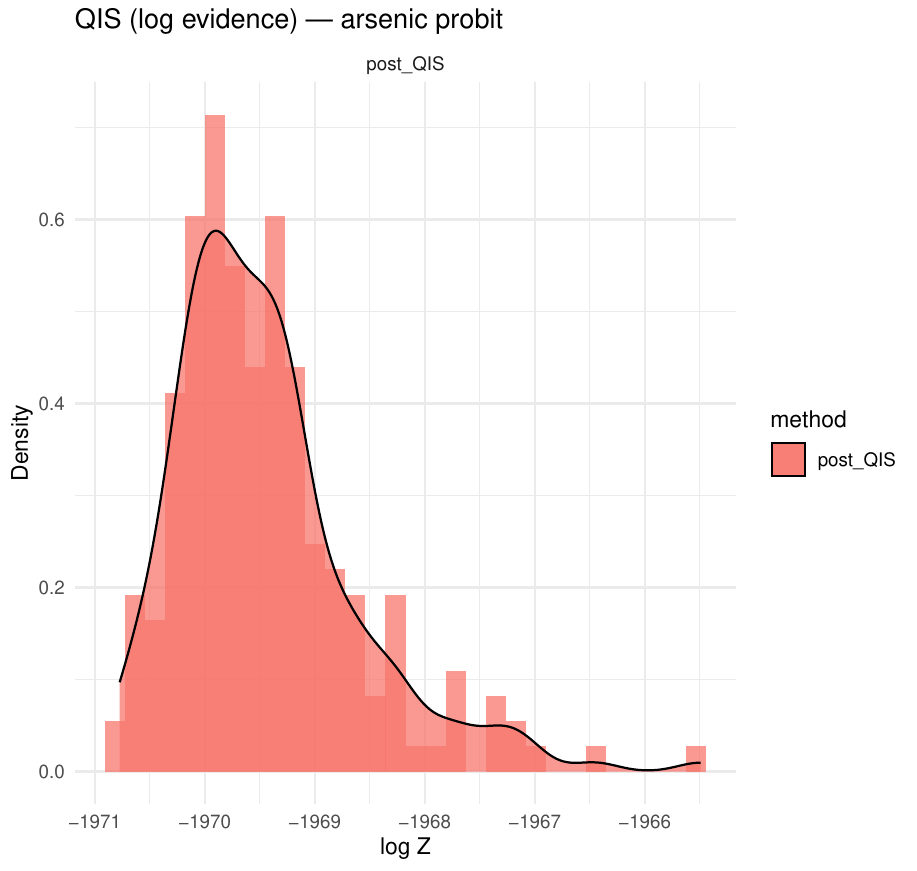}
\caption{Sampling distribution of $\log Z$ from QIS on the arsenic wells probit model (grid size $N=100$, $M=2000$ draws per replicate, $r=200$ replicates).}
\label{fig:qis-evidence}
\end{figure}
Using Quantile Importance Sampling with a Gaussian proposal $g=\mathcal N(\hat\theta,\hat\Sigma)$ obtained from a Laplace approximation at the posterior mode, and quantile integration of the importance weights $w(\theta)=\pi(\theta)L(\theta)/g(\theta)$ over a grid of $100$ live points with $2000$ draws per replicate and $200$ replicates, we obtain a sampling distribution for which the QIS mean and standard deviation of $\log Z$ are $-1969.446$ and $0.836406$ respectively; the resulting distribution is shown in Figure~\ref{fig:qis-evidence}. 

\section{Discussion} \label{sec:discussion}
Our main contribution in this paper is to take another look at the Riemann sum estimators for Monte Carlo integration in the context of nested sampling, a popular method for handling challenging high-dimensional integrals: aiding in model comparison and inference simultaneously. We show that under certain conditions, the Riemann sum estimator achieves a much faster convergence rate $O(n^{-4})$ compared to the existing results based on CLT. The other advantage of using the \textit{quantile reordering trick} or the simulated Lorenz curve is that it obviates the need to know an analytical form for the $\Lambda(s)$ or $Z(L(\x))$ function. However, there is one caveat to this result: if one does not know maximum of the likelihood ordinates, the last term can be $O(n^{-1})$, unfortunately. Finally, nested sampling may become inefficient in situations where likelihood is concentrated in regions with low prior probability \citet{chen2019improving}, and methods like Bayesian posterior repartitioning \citep{chen2019improving, chen2022bayesian} aim to handle this issue. \citet{chen2019improving} proposes defining an `effective prior' that keeps the unnormalized posterior constant while being broader than the original prior, and \citet{chen2022bayesian} uses a `power prior' approach by raising the original prior by a power in a Bayesian fashion. Since QIS depends only on quantiles of the likelihood ordinates calculated at prior draws, it is straightforward to apply the posterior repartitioning ideas to QIS and reap the efficiency benefits.

%%%%%%%%%%%%%%%%%%%%%%%%%%%%%%%%%%%%%%%%%%%%%%
%% Single Appendix:                         %%
%%%%%%%%%%%%%%%%%%%%%%%%%%%%%%%%%%%%%%%%%%%%%%
\begin{appendix}
\section{Appendix}

\subsection{Nested Sampling}\label{sec:nested}

The basic idea of nested sampling, due to \citet{skilling2006nested}, is to transform the problem of computing the evidence into a one-dimensional integration problem, where the integration variable is a decreasing sequence of likelihood values. \citet{polson2014vertical} presents an alternate representation of nested sampling from the vertical likelihood perspective. We will point the readers to the comprehensive review by \citet{ashton2022nested}. 

This is achieved by introducing a set of ``live points" that sample the likelihood space and are updated iteratively. At each iteration, the live point with the lowest likelihood is removed and replaced with a new point sampled from the likelihood space constrained by the likelihood of the removed point. This process continues until the evidence has been computed to a desired level of accuracy. Nested sampling has several advantages over other methods for computing the evidence, such as thermodynamic integration and importance sampling. It is generally more efficient and robust, and can handle multi-modal and high-dimensional distributions. We describe the algorithm below in Algorithm \ref{alg:nested_sampling}, following a similar terminology as \citet{ashton2022nested} and \citet{chen2022bayesian}. 

\begin{algorithm}[H]
\caption{Nested Sampling}
\label{alg:nested_sampling}
\begin{algorithmic}[1]
\Require $n_{live}$, the number of live points; $\mathcal{L}$, the likelihood function; $p(\btheta)$ prior.
\Ensure The evidence $Z$ and other quantities of interest.
\State Initialize $Z = 0$, and prior area/volume $X_0 = 1$.
\State Generate $n_{live}$ live points ${\btheta_1,\dots,\btheta_n} \sim p(\btheta)$.
\State Calculate the likelihood $\mathcal{L}_i=\mathcal{L}(\btheta_i)$, $i = 1, 2, \ldots, n_{live}$ for each live points.
\Repeat
\State{Identify the live point $\btheta_{i^*}$ with the lowest likelihood, call it $\mathcal{L}_{i^*}$.}
\State {Remove $\btheta_{i^*}$ from the set of live points.}
 \State {Sample a new point $\btheta_{\mathrm{new}}$ from $p(\btheta)$ subject to the constraint $\mathcal{L}> \mathcal{L}_{i^*}$.}
% \State {Let the new prior volume be $X_i = \exp(-i/n_{live})$, and the weight $\omega_i = \frac{X_{i-1}-X_{i+1}}{2}$.}
\State Calculate the contraction factor $\Delta X_i$ enclosed by the likelihood contour, i.e., 
\State $\Delta X_i = (1-t)X_i$, where $t = \e^{-1/n_{live}}$ and $X_{i} = t X_{i-1}$.
\State {Update the evidence: $Z = Z + \mathcal{L}_{i^*} \Delta X_i$.}
\State {$i \to i +1$.}
\Until {stopping criterion is met, e.g. $\max {\mathcal{L} (\btheta) } X_i > {\rm e}^{tol} Z$.}
\State Adjust for the remainder of evidence $Z = Z + \frac{1}{n_{live}} \sum_{n=1}^{n_{live}}\mathcal{L}(\btheta_n) X_I$.
\Return Evidence $Z$ and importance weights $p_i = \mathcal{L}_{i^*} \omega_i/Z, \; i=1, \ldots, I$.
\end{algorithmic}
\end{algorithm}

Note that this algorithm assumes a single likelihood peak, and may need to be modified for multimodal distributions. Additionally, the stopping criterion may vary depending on the specific problem being solved.

% \subsection{Proof of Lemma 1 as in \citep{yakowitz1978weighted}} 
% \begin{proof}
% From Wilks [11, 8.7.5 and 8.7.7], we have that the $Z_i$'s are jointly distributed according to the Dirichlet law, and that the marginal univariate and bivariate probability density functions are given, for $i = 1, 2, ... , n+1$ by
% $$
% f_{Z_i}(z) = \frac{\Gamma(n+1)}{\Gamma(n)} (1-z)^{n-1}, \qquad 0 \leq z \leq 1,
% $$
% and, for $i \neq j$, by
% $$
% f_{Z_i, Z_j}(z_i, z_j) = \frac{\Gamma(n+1)}{\Gamma(n-1)} (1-z_i-z_j)^{(n-2)}, \qquad z_i, z_j \geq 0, \qquad z_i+z_j \leq 1.
% $$
% Thus
% $$
% E[Z_i^3Z_j^3] = \int_0^1 \int_0 z_i^3 z_j^3 \frac{\Gamma(n+1)}{\Gamma(n-1)} (1-z_i-z_j)^{(n-2)} dz_idz_j.
% $$
% Note that since 
% $$
% \frac{\Gamma(n+7)}{\Gamma(4)\Gamma(4)\Gamma(n-1)} z_i^3 z_j^3 (1-z_i-z_j)^{(n-2)}
% $$
% is the density function for the Dirichlet variable with parameter $(4, 4; n-1)$  and consequently integrates to 1, we have that
% \begin{eqnarray*}
% E[Z_i^3Z_j^3] &=& \Gamma(4)^2\Gamma(n+1)/\Gamma(n+7)\\
% &=& (3!)^2n!/(n+6)!.
% \end{eqnarray*}
% The proof that $E[Z_i^6] = 6!n!/(n+6)!$ proceeds exactly as above.
% \end{proof}
\end{appendix}

%%%%%%%%%%%%%%%%%%%%%%%%%%%%%%%%%%%%%%%%%%%%%%
%% Multiple Appendixes:                     %%
%%%%%%%%%%%%%%%%%%%%%%%%%%%%%%%%%%%%%%%%%%%%%%
%\begin{appendix}
%\section{???}
%
%\section{???}
%
%\end{appendix}

%%%%%%%%%%%%%%%%%%%%%%%%%%%%%%%%%%%%%%%%%%%%%%
%% Support information, if any,             %%
%% should be provided in the                %%
%% Acknowledgements section.                %%
%%%%%%%%%%%%%%%%%%%%%%%%%%%%%%%%%%%%%%%%%%%%%%
\begin{acks}[Acknowledgments]
We thank Dr. Jianeng Xu for help with an earlier version of this manuscript and Dr. Sunanda Patra for helpful discussions on Monte Carlo methods.
\end{acks}

%%%%%%%%%%%%%%%%%%%%%%%%%%%%%%%%%%%%%%%%%%%%%%
%% Funding information, if any,             %%
%% should be provided in the                %%
%% funding section.                         %%
%%%%%%%%%%%%%%%%%%%%%%%%%%%%%%%%%%%%%%%%%%%%%%
\begin{funding}
Dr. Datta gratefully acknowledges support from the National Science Foundation (NSF CAREER Award DMS-2443282).
\end{funding}

%%%%%%%%%%%%%%%%%%%%%%%%%%%%%%%%%%%%%%%%%%%%%%
%% Supplementary Material, including data   %%
%% sets and code, should be provided in     %%
%% {supplement} environment with title      %%
%% and short description. It cannot be      %%
%% available exclusively as external link.  %%
%% All Supplementary Material must be       %%
%% available to the reader on Project       %%
%% Euclid with the published article.       %%
%%%%%%%%%%%%%%%%%%%%%%%%%%%%%%%%%%%%%%%%%%%%%%
%\begin{supplement}
%\stitle{???}
%\sdescription{???.}
%\end{supplement}

%%  imsart-nameyear.bst  will be used to                   %%
%%  MR numbers will be added by VTeX.                      %%
%%                                                         %%
%%  Use \cite{...} to cite references in text.             %%
%%                                                         %%
%%%%%%%%%%%%%%%%%%%%%%%%%%%%%%%%%%%%%%%%%%%%%%%%%%%%%%%%%%%%%

%% if your bibliography is in bibtex format, uncomment commands:
\bibliographystyle{imsart-nameyear} % Style BST file
\bibliography{references, sims_references}    % Bibliography file (usually '*.bib')

%% or include bibliography directly:
% \begin{thebibliography}{}
% \bibitem[\protect\citeauthoryear{???}{???}]{b1}
% \end{thebibliography}

\end{document}